\documentclass[showpacs,twocolumn,preprintnumbers,amsmath,amssymb,floatfix,prd]{revtex4}
\usepackage{graphicx}
\usepackage{dcolumn}
\usepackage{bm}
\newcommand{\be}{\begin{equation}}

\newcommand{\ee}{\end{equation}}
\newcommand{\bea}{\begin{eqnarray}}
\newcommand{\eea}{\end{eqnarray}}

\newcommand{\nn}{\nonumber}

\newcommand{\besplit}{\begin{split}}
\newcommand{\esplit}{\end{split}}

\begin{document}

\title{Cooling of a Compact Star with a LOFF Matter Core
}
\date{\today}
\author{Roberto~Anglani}\email{roberto.anglani@ba.infn.it}
\affiliation{Universit\`a di Bari, I-70126 Bari, Italia}
\affiliation{I.N.F.N., Sezione di Bari, I-70126 Bari, Italia}

\preprint{BARI-TH/06-553}
\begin{abstract}
Specific heat and neutrino emissivity due to direct URCA processes
for quark matter in the color superconductive
Larkin-Ovchinnikov-Fulde-Ferrell (LOFF) phase of
Quantum-Chromodynamics have been evaluated. The cooling rate of
simplified models of compact stars with a LOFF matter core is
estimated.
\end{abstract}

 \maketitle

Emission of $\beta$-neutrino due to direct URCA processes is the
dominant cooling mechanism for an aged compact star
\cite{ShapiroTeukolsky}. In the core of a neutron star, where
hadronic density is large enough to produce \cite{Collins}
deconfined quark matter, direct URCA processes can take place
\cite{Iwa}. At density and temperature relevant for aged pulsars,
quark matter could be in one of the possible color superconducting
(CSC) phases \cite{Alford:1997zt}. The ground state of a color and
electrically neutral matter of deconfined quarks in weak equilibrium
is still a matter of debate \cite{Alford:2006fw}. A phase named LOFF
which is energetically favored with respect to the normal quark
matter and gapless phases \cite{gapless} in a wide range of
densities, has attracted theoretical attention \cite{LOFF}. This
phase turns out to be also chromomagnetically stable
\cite{Ciminale:2006sm}. The LOFF pairing is characterized by a non
vanishing total momentum of the pair. The one-plane wave structure
ansatz, defined by
\begin{equation}\label{ansatz}
\langle\psi_{\alpha i}(x)C \gamma_5
\psi_{\beta j}(x) \rangle \propto
\sum_{I=1}^3 \Delta_I\, e^{2i{\bf
q}_I\cdot{\bf r}}\epsilon_{\alpha\beta
I}\epsilon _{i j I}
\end{equation}
($i,j=1,2,3$ flavor indices, $\alpha,\beta=1,2,3$ color indices) has
been considered in Ref. \cite{Nicola} and found energetically
favored with respect to the gapless and the unpaired phases in a
certain range of values of $\delta\mu$. In Eq.~(\ref{ansatz}),
$2\,{\bf q}_I$ represents the momentum of the Cooper pair and the
gap parameters $\Delta_{1}$, $\Delta_{2}$, $\Delta_{3}$ describe
respectively $d-s$, $u-s$ and $u-d$ pairing. For sufficiently large
$\mu$ the energetically favored phase is characterized by
$\Delta_1=0$, $\Delta_2=\Delta_3$ and ${\bf q}_2 = {\bf q}_3$.

The presence of quark matter in the LOFF state in the core of a
pulsar will affect the neutrino emissivity and consequently the
cooling process. Neutrino emissivity is defined as the energy loss
by $\beta$-decay per volume unit per time unit. We have discussed
these effects by comparing various models of stars similarly to the
analysis performed in Ref.~\cite{Alford:2004zr} for the gCFL phase
(for similar calculations see Ref.~\cite{Sedrakian}). We have used
\cite{Anglani:2006br} a simplified approach based on the study of
three different toy models of stars. The first model (denoted as I)
is a star consisting of noninteracting ``nuclear" matter (neutrons,
protons and electrons) with mass $M=1.4M_\odot$, radius $R=12$ km
and uniform density $n=1.5\,n_0$, where $n_0 = 0.16$ fm$^{-3}$ is
the nuclear equilibrium density. The nuclear matter is assumed to be
electrically neutral and in beta equilibrium. The second model (II)
is a star containing a core of radius $R_1=5$ km of neutral unpaired
quark matter at $\mu=500$ MeV, with a mantle of noninteracting
nuclear matter with uniform density $n$. Solution of the
Tolman-Oppenheimer-Volkov equations gives a mass-radius relation so
that a mass $M=1.4\,M_{\odot}$ corresponds to a star radius $R_2=10
$ km. The model III is represented by a compact star containing a
core of electric and color neutral three flavors quark matter in the
LOFF phase, with $\mu=500$ MeV and $m_s^2/\mu=140$ MeV.

\begin{figure}[t]
\begin{center}
\includegraphics[width=8cm]{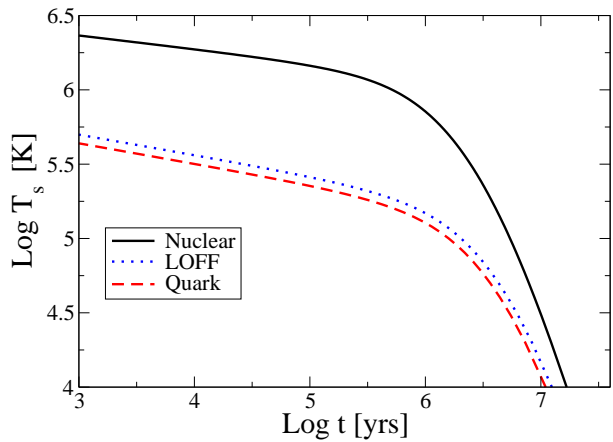}
\end{center}
\caption{\footnotesize{(Color online) Surface
temperature $T_s$, in Kelvin, as a function
of time, in years, for the three toy models
of pulsars described. Solid black curve
refers to a neutron star formed by nuclear
matter with uniform density $n=0.24$
fm$^{-3}$ and radius $R=12$ Km (model I);
dashed line (red online) refers to a star
with $R_2=10$ km, having a mantle of nuclear
matter and a core of radius $R_1=5$ Km of
unpaired quark matter, interacting {\it via}
gluon exchange (model II); dotted curve (blue
online) refers to a star like model II, but
in the core there is quark matter in the LOFF
state; see \cite{Anglani:2006br} for more
details. All stars have $M=1.4\,M_\odot$.}}
\label{coolsurf}
\end{figure}

The main processes of cooling are dominated by neutrino emission in
the early stage of the lifetime of the pulsar and by photon emission
at later ages. The cooling rate is governed by the following
differential equation:
\begin{eqnarray} \nn\frac{dT}{dt} &=& - \frac{
L_\nu+L_\gamma}{V_{nm}c_V^{nm} + V_{qm}c_V^{qm}}\\ \quad&=& -
\frac{V_{nm}\varepsilon_\nu^{nm} + V_{qm}\varepsilon_\nu^{qm} +
L_\gamma} {V_{nm}c_V^{nm} + V_{qm}c_V^{qm}}~. \label{ANGLANI1}
\end{eqnarray} Here $T$ is the inner temperature at time $t$;
$L_\nu$ and $L_\gamma$ are neutrino and photon luminosities, i.e.
emissivity by the corresponding volume. The superscripts $nm$ and
$qm$ refer, respectively, to nuclear matter and quark matter
including the superconductive phase; $c_V^{nm}$ and $c_V^{qm}$
denote specific heats of the two forms of hadronic matter. Eq.
(\ref{ANGLANI1}) is solved imposing a given temperature $T_0$ at a
fixed early time $t_0$ (we use $T_0\to\infty$ for  $t_0\to 0$). More
detailed results are reported in Ref.~\cite{Anglani:2006br}. In Fig.
\ref{coolsurf} we can see the star surface temperature as a function
of time. Solid line (black online) is for model I; dashed curve (red
online) refers to model II; the dotted line (blue online) is for
model III and it is obtained for the following values of the
parameters: $\mu=500$ MeV, $m_s^2/\mu=140$ MeV, $\Delta_1=0,\,
\Delta_2=\Delta_3\simeq 6$ MeV, where $m_s$ is the strange quark
mass. Let us observe that the neutrino emissivity decreases at
increasing values of $m_s$. This is due to the fact that $\Delta$
decreases as one approaches the second order phase transition to the
normal state. At the same time the quark matter tends to a normal
Fermi liquid state, but in this case the description of Iwamoto
which includes Fermi liquid effects and one gluon exchange, must be
adopted. For unpaired quark matter we use $\alpha_s\simeq 1$, the
value corresponding to $\mu=500$ MeV and $\Lambda_{\rm QCD}=250$
MeV. The use of perturbative QCD at such small momentum scales is
however questionable. Therefore the results for model II should be
considered with some caution and the curve is plotted only to allow
a comparison with the other models. In any case it is important to
remark that the apparent similarity between the LOFF curve and the
unpaired quark curve depends on the fact that the LOFF phase is
gapless. This yields a parametric dependence on temperature
analogous to that of the unpaired quark matter: $c_V\sim T$ and
$\varepsilon_\nu\sim T^6$. However the similarity between the curves
of models II and III should be considered accidental because
emissivity of unpaired quark matter depends on the value we assumed
for the strong coupling constant.
From Fig.~\ref{coolsurf} we see that the temperature of stars with a
LOFF core drops faster than ordinary neutron stars. Some interesting
phenomenological consequences could exist (for compilation of data
and comparison between theoretical models and data see, e.g.
\cite{blaschke}). Slow cooling is typical of stars made of ordinary
nuclear matter or of stars with a CSC quark core in a gapped phase
(e.g. CFL). If a careful comparison with the data could allow to
rule out slow cooling for star masses, in the range we have
considered, this would favor either the presence of condensed mesons
or quark matter in one of the gapless states in the core. Among them
the LOFF state is favored since we know that other phases with
homogeneous gap parameters such as, e.g. the gCFL or the g2SC phase
are instable \cite{instability}. Let us finally observe that our
results should be considered as preliminary, since the simple ansatz
(\ref{ansatz}) should be substituted by a more complex structure as
in \cite{cubex}. Notwithstanding this question, our conclusions
should remain valid, at least qualitatively, also for more complex
crystalline patterns of the LOFF condensate. The existence of
gapless points and of blocking regions in momentum space
characterize the scaling laws for neutrino emissivity and specific
heat. Since these properties are typical of the LOFF state,
independently of the detailed form of the condensate, a rapid
cooling should be appropriate not only for the simple ansatz we
assumed, but, more generally, for any LOFF condensate. \vskip0.3cm I
would like to thank D. Blaschke and BLTP-JINR for the pleasant
organization of DM2006 school and for kind hospitality; M. Ruggieri,
G. Nardulli and M. Mannarelli for the fruitful collaboration which
has yielded the work \cite{Anglani:2006br} whose results underlie
the present paper; N. Ippolito, I. Parenti, M. Buballa and H.
Grigorian for useful discussions during the school.

\end{document}